\documentclass{aa}

\usepackage{amssymb,amsmath}
\usepackage{epstopdf}
\bibpunct{(}{)}{;}{a}{}{,} 
\usepackage{color, xcolor}
\usepackage{hyperref}
\usepackage{fancyvrb}
\usepackage{graphicx}
\usepackage{multirow}
\usepackage{appendix}
\usepackage{color}
\usepackage{natbib}
\usepackage{ulem}
\usepackage{subcaption, caption}
\usepackage{tabularray}

\newcommand{\be}{\begin{equation}}
\newcommand{\ee}{\end{equation}}
\newcommand{\bea}{\begin{eqnarray}}
\newcommand{\eea}{\end{eqnarray}}
\newcommand{\stkout}[1]{\ifmmode\text{\sout{\ensuremath{#1}}}\else\sout{#1}\fi}

\begin{document}

\title{Bayesian analysis of the shear modulus in the neutron-star crust}

\author{T. Diverr\`es\inst{1}, G. Montefusco\inst{2}, A. F. Fantina\inst{1}, F. Gulminelli\inst{2,3}}

\institute{Grand Acc\'el\'erateur National d'Ions Lourds (GANIL), CEA/DRF - CNRS/IN2P3, Boulevard Henri Becquerel, 14076 Caen, France \\
\email{theau.diverres@ganil.fr, anthea.fantina@ganil.fr}
\and Université de Caen Normandie, ENSICAEN, CNRS/IN2P3, LPC Caen UMR6534, F-14000 Caen, France 
\and Institut Universitaire de France (IUF), France
}

\titlerunning{Bayesian analysis of the shear modulus in the neutron-star crust}
\authorrunning{Diverr\`es et al.}

\abstract
   {The elastic properties of the neutron-star crust are important for the calculations of crustal modes. In particular, the ability of the crust  to  support  shear  stresses  has  been connected to observations of quasi-periodic oscillations and to crust deformations potentially emitting gravitational waves.
   }
    {In this work, we assess the uncertainties in the shear modulus and shear speed in the neutron-star outer and inner crust. 
    }
   {To this aim, we performed a Bayesian analysis of the shear properties of the neutron-star crust at zero temperature starting from both a non-informative and a nuclear-physics-informed prior. For the treatment of inhomogeneous matter in the crust, we relied on the one-component plasma approximation, with a (semi-)classical treatment of the ions. 
   } 
   {We show that the use of a nuclear-physics-informed prior has a non-negligible impact on the prediction of the elastic properties of the crust. The frequency of the fundamental torsional crustal modes we obtained is compatible with the low-frequency range of observed quasi-periodic oscillations, with our estimates lying in the interval $\approx 20 - 50$~Hz.
   }
   {Although the different considered priors lead to compatible results, the inclusion of nuclear-physics experimental information in the prior considerably reduces the uncertainties in the prediction of the elastic properties of the crust, potentially constraining the predicted frequency of the crustal modes.
   } 


\maketitle
\nolinenumbers

\section{Introduction}
\label{sec:intro}

The structure of a cold non-accreting neutron star (NS) is expected to consist of different regions: an outer crust (corresponding to the first few hundred metres under the NS surface), made of fully ionised atoms immersed in a charge-compensating electron background, an inner crust (from the so-called neutron drip at about $2 \times 10^{-4}$~fm$^{-3}$ up to the crust-core transition at about $0.5 n_{\rm sat}$, the saturation density being $n_{\rm sat} \approx 0.16$~fm$^{-3}$), where neutron-proton clusters co-exist with a background of unbound neutrons and electrons, and a core, made of homogeneous matter where non-nucleonic degrees of freedom might also be present (see e.g. \citet{Haensel2007, Oertel2017, Burgio2018, Blaschke2018, Raduta2022} for a review).
In this so-called `cold catalysed matter' hypothesis, which states that the NS composition can be determined in the ground state at zero temperature, the crust is expected to be stratified in pure layers, each of which consists of a one-component Coulomb crystal (except possibly at the boundaries between adjacent layers; see \citet{Chamel2016} for a discussion on the outer crust).
Therefore, in this scenario $-$ and below the crystallisation temperature overall \citep{Haensel2007, Medin2010, Fantina2020, Carreau2020} $-$ nuclei in the crust are expected to be crystallised and form a body-centred-cubic (bcc) lattice; however, we do note that different geometries (e.g. a face-centred-cubic lattice) have been found to be thermodynamically favoured in some cases (see e.g. \citet{Kozhberov2021}).
Therefore, the crust is expected to behave as a solid, allowing for elasticity.

The computation of the elastic properties of the NS crust has recently regained attention since the ability of the crust to support shear stresses has been connected to observations.
Crustal elasticity has indeed been suggested to explain quasi-periodic oscillations (QPOs) following giant flares of magnetars (see e.g. \citet{Mcdermott1988, Gabler2018, Sotani2018, Kozhberov2020, Sotani2024}).
In addition, the shear (and bulk) modulus are relevant for modelling crust deformations, such as so-called `mountains' (i.e. asymmetric deformations in the crustal mass distribution) that can lead to emission of gravitational waves (see e.g. \citet{Haskell2006, Gearheart2011, Gittins2020, Kerin2022}).
In particular, it has been pointed out that the overall scaling of deformation is set by the shear speed \citep{Giliberti2020}.

Several works in the literature, based on different approaches and approximations, have been devoted to the determination of the elastic properties (and, in particular, of the shear modulus) of the crust, not only in the context of NSs, but also of white dwarfs.
Usually, computations are carried out with the assumption that the crust is an isotropic elastic material (although the anisotropy of the crustal elastic properties has been studied; see e.g. \citet{Baiko2024, Morales2024}) and that the lattice is characterised by a cubic symmetry (see e.g. \citet{Marder2000} for a textbook discussion on elasticity).
Investigations have been performed for monocrystalline structure (i.e. where all crystals, also referred to as `grains', have the same composition and orientation) and polycrystalline structure of the crust (i.e. where  the crystallites have the same composition but different random orientations). 
These studies include earlier works by \citet{BaymPines1971} and those of \citet{Ogata1990} and \citet{Strohmayer1991}, who computed the effective shear modulus for (bcc) Coulomb solids (and `quenched' solids too) using Monte-Carlo simulations, while more recent calculations have been presented for example in \citet{Horowitz2008, Chugunov2010, Kobyakov2015, Chugunov2021}.
Subsequent estimates have accounted for additional effects in the shear modulus, such as electron screening, quantum motion of nuclei, or nuclear finite size (see e.g. \citet{Horowitz2008,Hoffman2012, Baiko2012, Baiko2015, Kozhberov2022, Sotani2022, Zemlyakov2023a}), while other studies have also considered multi-component lattices (see e.g. \citet{Kozhberov2019, Chugunov2022}).
Moreover, the shear modulus has been computed for non-spherical configurations, the so-called `pasta phases' that can appear at the bottom of the inner crust, the inclusion of non-spherical structures generally yielding a reduction of the shear modulus (see e.g. \citet{Pethick1998, Caplan2018prl, Pethick2020, Xia2023, Zemlyakov2023b}).

In this work, we present a Bayesian analysis of the shear properties of the NS crust at zero temperature to assess the present uncertainties in the shear modulus and speed, needed for the calculations of the fundamental crustal modes. In particular, we address the uncertainties coming from our incomplete knowledge of the nuclear energy functional, specifically by broadly varying the model parameters in our prior.
To this aim, we used the estimate of the shear modulus given in \citet{Strohmayer1991}, as this expression has been extensively used in the literature to study torsional oscillations and interface modes (see e.g. \citet{Piro2005, Steiner2009, Passamonti2016, Tews2017, Sotani2013}, and also \citet{Neill2023, Sotanietal2024PRD} for very recent works).
We also included the correction to the point-like-nucleus expression of \citet{Strohmayer1991} as derived in \citet{Zemlyakov2023a}, but we neglected the impact of non-spherical configurations in our first computation of the shear modulus presented here. 

The paper is organised as follows: we describe the theoretical framework used to model the crust in Sect.~\ref{sec:method}. In particular, we present the modelling of the inhomogeneous matter in Sect.~\ref{sec:cldm} and the expressions for the elastic properties used in this study in Sect.~\ref{sec:shear}. The results are presented in Sect.~\ref{sec:results}: the set-up for the Bayesian analyses is described in Sect.~\ref{sec:bayes} (with additional details given in Appendix~\ref{app:bayes}), while the results on the shear properties are given in Sect.~\ref{sec:res-shear} and an estimate of the fundamental-mode frequency is provided in Sect.~\ref{sec:res-freq} (with complementary results provided in Appendix~\ref{app:median}). Finally, our conclusions are reported in Sect.~\ref{sec:conclusions}.

\section{Model of the neutron-star crust}
\label{sec:method}

In this work we consider NSs at zero temperature in the one-component plasma approximation \citep{Haensel2007, Chamel2008}. 
According to the latter hypothesis, each layer of the crust is composed of identical Wigner-Seitz (WS) cells, each with a volume $V_{\rm WS}$, with an ion with proton number, $Z$, and mass number, $A$, at the centre, surrounded by a uniform electron gas (and in the inner crust, unbound neutrons as well).
In view of the uncertainties pertaining to the calculation of the shear modulus in the so-called `pasta phases' that are expected to appear at the bottom of the inner crust, we considered only spherical clusters (see e.g. \citet{Pethick1998, Caplan2018prl, Pethick2020, Xia2023, Zemlyakov2023b} for a calculation of elastic properties in non-spherical configurations and \citet{Dinh2021a} for a discussion on pasta phases within the same approach used here).

\subsection{Inhomogeneous matter in the crust}
\label{sec:cldm}

Following the seminal work by \citet{Baym1971}, the equilibrium configuration of inhomogeneous (catalysed) matter in the NS crust is obtained within a compressible liquid-drop model (CLDM) approach in the one-component plasma approximation. Despite its simpler form, CLDM results agree well with more microscopic approaches and it offers the advantage of being computationally faster with respect to more microscopic calculations (see e.g. \citet{Carreau2020,Grams2022,Grams2025}). In this framework, the energy density in each WS cell is written as\footnote{We use the uppercase $E$, the lowercase $e$, and $\mathcal{E}$ to denote the total energy per WS cell, the energy per nucleon, and the energy per unit volume, respectively.}
\begin{equation}
    \mathcal{E} = \mathcal{E}_e +  \mathcal{E}_{\rm g} (1-u) + \frac{E_i}{V_{\rm WS}} \ 
    \label{eq:energy-density},
\end{equation}
where $\mathcal{E}_e(n_e)$ is the energy density of the electron gas at a density $n_e$ (the electron mass is also included),
\begin{equation}
\mathcal{E}_e=\frac{(m_e c^2)^4}{8 \pi^2 (\hbar c)^3}\left [ x_r(2x_r^2+1)\sqrt{x_r^2+1}- \ln \left( x_r + \sqrt{x_r^2+1} \right) \right ] \;,
\label{eq:e_el}
\end{equation}
with $x_r= \hbar c (3\pi^2 n_e)^{1/3}/(m_e c^2)$ being the relativity parameter, $\hbar$ the Dirac-Planck constant, $c$ the speed of light, and $m_e$ the electron mass (see e.g. \citet{Weiss2004}), $\mathcal{E}_{\rm g} = \mathcal{E}_{\rm nuc}(n_{\rm g}, \delta_{\rm g}) + n_{\rm g} m_n c^2$ is the neutron-gas energy density (including the rest mass $m_n$ of neutrons) at a baryonic density $n_{\rm g}$ and isospin asymmetry $\delta_{\rm g} = 1$ (no protons are present in the gas at $T=0$), and $u=A/(n_i V_{\rm WS})$ is the volume fraction of the cluster with internal density $n_i$. The third term on the right hand side of Eq.~(\ref{eq:energy-density}), $- \mathcal{E}_{\rm g} u$, accounts for the excluded volume. In the outer crust, $n_{\rm g}$ (and, thus, $\mathcal{E}_{\rm g}$) is set to zero.
In the last term in Eq.~(\ref{eq:energy-density}), the cluster energy per Wigner-Seitz cell is expressed as
\begin{equation}
    E_i = M_i c^2 + E_{\rm bulk} + E_{\rm Coul} + E_{\rm surf +  curv} \ 
    \label{eq:Ei},
\end{equation}
where $M_i = (A-Z)m_n + Zm_p$ is the total bare mass of the cluster ($m_p$ being the proton mass), $E_{\rm bulk} = A e_{\rm nuc}(n_i, 1-2Z/A)$ is the cluster bulk energy, and $E_{\rm Coul} + E_{\rm surf + curv} = V_{\rm WS} (\mathcal{E}_{\rm Coul} + \mathcal{E}_{\rm surf} + \mathcal{E}_{\rm curv})$ accounts for the total interface energy, that is, the Coulomb interaction between the nucleus and the electron gas (including the proton-proton, proton-electron, and electron-electron contributions) as well as the residual interface interaction between the nucleus and the surrounding dilute nuclear-matter medium.
For the bulk energy density of the ions, $\mathcal{E}_{\rm bulk} = e_{\rm nuc}n_i$, and that of the neutron gas, we used the same functional expression, that is, the meta-model approach of \citet{Margueron2018} (see Eq.~\eqref{eq:meta}).
As we considered only spherical clusters, the Coulomb energy density \citep{Ravenhall83} reads
\begin{equation}
    \mathcal{E}_{\rm Coul}  = \frac{2}{5}\pi (e n_i r_N)^2 u \left(\frac{1-I}{2}\right)^2 \left[ u+ 2  \left( 1- \frac{3}{2}u^{1/3} \right) \right] \ , 
    \label{eq:Fcoul}
\end{equation}
with $e$ being the elementary charge, $I = 1-2Z/A$, and $r_N = (3A/(4 \pi n_i))^{1/3}$. For the surface and curvature contributions, we employed the same expression as in \citet{Maruyama2005} and \citet{Newton2013}, namely,
\begin{equation}
\mathcal{E}_{\rm {surf}} + \mathcal{E}_{\rm {curv}} =\frac{3u}{r_N} \left( \sigma_{\rm s}(I) +\frac{2\sigma_{\rm c}(I)}{r_N} \right) \ , 
\label{eq:interface}   
\end{equation}
where $\sigma_{\rm s}(I)$ and $\sigma_{\rm c}(I)$ are the surface and curvature tensions \citep{Ravenhall1983},
\begin{eqnarray}
\sigma_{\rm s}(I) &=& \sigma_0 \frac{2^{p+1} + b_{\rm s}}{y_p^{-p} + b_{\rm s} + (1-y_p)^{-p}} \ , \label{eq:sigma0_1} \\
\sigma_{\rm c} (I) &=& 5.5 \sigma_{\rm s}(I) \frac{\sigma_{0, {\rm c}}}{\sigma_0} (\beta -y_p) \ ,
\label{eq:sigma0_2}
\end{eqnarray}
where $y_p = (1-I)/2$, and the surface parameters $(\sigma_0, \sigma_{0, {\rm c}}, b_{\rm s}, \beta)$ were optimised for each set of bulk parameters and effective mass to reproduce the experimental nuclear masses in the 2020 atomic mass evaluation (AME) table \citep{ame2020}, while we set $p=3$ \citep{Carreau2019b}. The equation of state and composition of the crust were then obtained by variationally minimising the energy density of the Wigner-Seitz cell with $(A, I, n_i, n_e, n_{\rm g})$ as variational variables, under the constraint of baryon number conservation,
\begin{equation}
    n_B = \frac{A}{V_{\rm WS}} + n_{\rm g} (1-u) \ ,
    \label{eq:bar-cons}
\end{equation}
and charge neutrality holding in every cell \citep{Carreau2019, Dinh2021a, Davis2024}.

For the nuclear functional, $e_{\rm nuc}$, we followed the approach of \citet{Margueron2018}, decomposing the energy contributions in a `kinetic' and a `potential' part via
\begin{equation}
e_{\rm nuc}(n_B,\delta) = t_{\rm FG}^\star(n_B,\delta) + v_{\rm MM}(n_B, \delta) \ ,
\label{eq:meta}
\end{equation}
with $\delta=(n_n-n_p)/(n_n+n_p)$, $n_n$ ($n_p$) being the neutron (proton) density.
The kinetic term, $t_{\rm FG}^\star$, includes the dominant deviation to the parabolic approximation\footnote{Indeed, because of this term, Eq.~\eqref{eq:meta} is not a purely parabolic approximation for the isospin dependence.} as well as the effective mass contribution,
\begin{eqnarray}
t_{\rm FG}^\star(n_B,\delta) &=& \frac{t_{\rm FG, sat}}{2}\left(\frac{n_B}{n_{\rm sat}}\right)^{2/3} 
\bigg[ \left( 1+\kappa_{\rm sat}\frac{n_B}{n_{\rm sat}} \right) f_1(\delta) \nonumber \\
&& + \kappa_{\rm sym}\frac{n_B}{n_{\rm sat}}f_2(\delta)\bigg] ,
\label{eq:effmassms2}
\end{eqnarray}
where $t_{\rm FG, sat} = 3 \hbar^2 c^2 \left(3\pi^{2}/2\right)^{2/3}n_{\rm sat}^{2/3}/(10 m c^2) $ is the kinetic energy per nucleon of symmetric nuclear matter at the saturation density $n_{\rm sat}$, with $m=(m_n+m_p)/2$. The functions, $f_i$, are defined as
\bea
f_1(\delta) &=& (1+\delta)^{5/3}+(1-\delta)^{5/3}, \ \\
f_2(\delta) &=& \delta \left[ (1+\delta)^{5/3}-(1-\delta)^{5/3} \right]. \
\eea
The terms $\kappa_{\rm sat}$ and $\kappa_{\rm sym}$ in Eq.~(\ref{eq:effmassms2}) are linked to the nucleon effective masses $m_q^\star$ ($q=n,p$ labelling neutrons and protons, respectively) via
\begin{eqnarray}
\kappa_{\rm sat}&=&\frac{m}{m_{\rm sat}^\star} - 1 \;\;  \hbox{ in symmetric matter ($\delta=0$)}  \;, \nonumber \\
\kappa_{\rm sym}&=&\frac 1 2 \left[ \frac{m}{m^\star_n} - \frac{m}{m^\star_p}  \right] \hbox{ in neutron matter ($\delta=1$)} \ ,
\end{eqnarray}
where $m_{\rm sat}^\star = m^\star(n_{\rm sat})$ is the Landau effective mass at saturation (see also the discussion in \citet{Margueron2018}).
For the potential part, $v_{\rm MM}$ in Eq.~\eqref{eq:meta}, we used the meta-model approach based on a Taylor expansion in $x = (n_B-n_{\rm sat})/(3 n_{\rm sat})$ up to order $N$ around the saturation point ($n_B = n_{\rm sat}$, $\delta = 0$) \citep{Margueron2018}:
\begin{equation}
v_{\rm MM} = \sum_{k = 0}^{N} \dfrac{1}{k!} \left(v_{k}^{\rm is} + v_{k}^{\rm iv} \delta^{2}  \right) x^{k} u_k(x) \ ,
\label{eq:vmm}
\end{equation}
where the function $u_k(x)$, expressed as
\begin{equation}
u_k(x)=1-(-3x)^{N+1-k} e^{-b n_B/n_{\rm sat}} \ ,
\label{eq:uk}
\end{equation}
ensures the correct zero-density limit, with $b$ determining the exponential decay of $u$ for $n_B \rightarrow 0$, and the parameters $v_k^{\rm is(iv)}$ are directly connected to the so-called isoscalar (isovector) nuclear empirical parameters (see \citet{Margueron2018} for details). 
When truncating the expansion at order $N=4$, the meta-model has been found to offer a very good reproduction of realistic functionals at low density (see \citet{Margueron2018} and \citet{Davis2024} for a discussion on the truncation order). 
Moreover, this parameterised energy functional can interpolate among existing energy-density functionals and possibly explore new density dependencies not yet proposed in the literature, allowing for an easy application to Bayesian studies, such as the one carried out in the present work.
Recently, a correction to the meta-model at very low densities based on ab initio calculations was proposed in \citet{Burrello2025}, but we defer the implementation of such protocol to a future work.
We also note that the same expression for the bulk energy per baryon, $e_{\rm nuc}$, was used for the ion (see Eq.~\eqref{eq:Ei}) and the neutron gas, allowing for a unified treatment of the nucleons `inside' and `outside' the cluster.

\subsection{Shear modulus and speed}
\label{sec:shear}

As mentioned in the introduction, different works have been devoted in the literature to study the shear modulus and derive analytical expressions to be applied in astrophysical applications. Here, we used the fit proposed in \citet{Strohmayer1991} based on Monte-Carlo calculations by \citet{Ogata1990},
\begin{equation}
    \mu = 0.1194\ \frac{n_N (Z e)^2}{a} \ ,
    \label{eq:shear-mod}
\end{equation}
with $n_N = 1/V_{\rm WS}$ being the ion density and $a=(4\pi n_N / 3)^{-1/3}$. This latter expression was widely employed in the literature to study, for example, torsional oscillations and interface modes (see e.g. \citet{Kruger2015,Passamonti2016,Kozhberov2020, Sotanietal2024PRD,Neill2023}).
We also included the correction factor for the ground-state crust proposed in \citet{Zemlyakov2023a} that accounts for the deviation from the expression for point-like nuclei. Equation~\eqref{eq:shear-mod} thus becomes
\begin{equation}
    \mu = 0.1194\ \frac{n_N (Z e)^2}{a} \left( 1 - \frac{u^{5/3}}{2 - 4 u^{1/3} + 3 u} \right) \ ,
    \label{eq:shear-mod-corr}
\end{equation}
with $u$ being the volume fraction of the cluster. Therefore, in our first estimate of the shear modulus, we neglected electron-screening corrections and the impact of pasta phases (see e.g. \citet{Chugunov2022, Zemlyakov2023b} for a discussion). 

From the shear modulus, we can calculate the shear speed \citep{Marder2000}, expressed as
\begin{equation}
    v_s = \sqrt{\frac{\mu}{\rho_{B,{\rm d}}}} \ ,
    \label{eq:shear-speed}
\end{equation}
with $\rho_{B,{\rm d}}$ being the `dynamical' mass density, that is, the mass density of nucleons moving with the lattice.
Because of superfluidity, which is expected to manifest itself in the NS crust, $\rho_{B,{\rm d}}$ does not coincide with the total mass density, $\rho_B = m n_B$; in fact, it is reduced, thereby increasing the shear speed \citep{Andersson2009, Sotani2013, Tews2017}.
The calculation of the mobility of superfluid neutrons in the NS crust is not a trivial task. Various works in the literature have been devoted to compute the so-called `entrainment' to quantify the fraction of unbound neutrons effectively co-moving with the lattice, due to Bragg scattering (see e.g. \citet{Chamel2012, Martin2016, Chamel2017, Kashiwaba2019, Sekizawa2022, Almirante2025}).
Although \citet{Chamel2012} found that a large fraction (up to $90\%$) of unbound neutrons in the NS crust could be entrained, more recent microscopic calculations \citep{Almirante2025} suggest that this effect may be significantly weaker, with an entrained proportion of the order of $10\%$ of the total number of neutrons.
The latter calculations are in fact in agreement with previous results by \citet{Martin2016} obtained using a superfluid hydrodynamics approach. The authors of that work also showed (see their Fig.~9) that the neutron superfluid fraction is adequately reproduced by the `free' neutrons in the e-cluster representation \citep{Papak2013}; namely, neutrons that are energetically unbound. Therefore, we considered this latter case as the most realistic, where the dynamical mass density is given by\footnote{Strictly speaking, the use of the mass density in the calculation of the shear speed is derived for the Newtonian case. In studies working in the framework of general relativity, often the shear speed is defined through the enthalpy density, that is, replacing $\rho_{B}$ with $\rho+P/c^2$, with $\rho$ being the mass-energy density (see e.g. \citet{Samuelsson2007} and their Eq.~(A13), and \citet{Sotani2013}). However, we checked that the use of the enthalpy density instead of the mass density induces differences in the shear speed of about a few percents ($\sim 2 \%$ at most for calculations using the BSk24 empirical parameters and considering superfluidity, see Eq.~\eqref{eq:rhobd}). Therefore, in this work, we employed the mass density $\rho_B$ in Eq.~\eqref{eq:shear-speed}.}
\begin{equation}
 \rho_{B,{\rm d}} = \rho_B - \rho_{B,{\rm g}} \ ,
 \label{eq:rhobd}
\end{equation}
with $\rho_{B,{\rm g}} = m n_{\rm g}$. Unless specified, all our calculations of the shear properties are performed under this hypothesis.

The shear speed can be also used to evaluate the frequency of the fundamental ($n=0$) torsional oscillation crust mode, denoted as $_lt_0$. The latter can be estimated through a plane-wave analysis of the crustal (axial) perturbation equations as \citep{Piro2005, Samuelsson2007}
\begin{equation}
    \omega_0 \approx \frac{e^{2 \nu} v_s^2 (l-1) (l+2)}{2 R R_{\rm cc}} \ ,
    \label{eq:omega}
\end{equation}
where $e^{2 \nu} = 1 - 2GM/(Rc^2)$, with $M$ and $R$ being the NS mass and radius, respectively, $G$ the gravitational constant, and $R_{\rm cc}$ the radius at the crust-core transition. As discussed in \citet{Samuelsson2007}, the approximation in Eq.~\eqref{eq:omega} is based on the assumption $\omega_0 \gg e^{2 \nu} v_s^2 (l-1) (l+2)/r^2$, which is not necessarily satisfied throughout the crust. Moreover, to compute $\omega_0$, we can either assume an average value of $v_s$ (for example, a density-weighted average, as in \citet{Neill2023}) or define its value at a given density in the crust; for instance, some works in the literature considering pasta phases take $v_s$ either at the transition from spherical to non-spherical configurations, or at the crust--core transition (see e.g. \citet{Gearheart2011, Parmar2022}).
In this work, we chose to calculate an average value of the shear speed, to avoid an overestimation (or underestimation) of $v_s$ at a specific density-point in the crust (see the discussion in Sect.~\ref{sec:res-freq}).
Although it remains a rough estimate, this result provides a useful order-of-magnitude evaluation avoiding the full resolution of the perturbation equations and allowing comparison with existing computations in the literature (see e.g. \citet{Tews2017, Gearheart2011, Parmar2022, Sotanietal2024PRD}).

\section{Numerical results}
\label{sec:results}

In this section, we present estimates of the uncertainties in the elastic properties related to the shear modulus, described in Sect.~\ref{sec:shear}, in the NS crust, obtained by performing Bayesian analyses.
We describe the set-up for our analysis in Sect.~\ref{sec:bayes} and show our results in Sects.~\ref{sec:res-shear} and \ref{sec:res-freq}.

\subsection{Bayesian analysis: Set-up}
\label{sec:bayes}

Each model is characterised by 17 parameters: the nuclear empirical parameters $\{ n_{\rm sat}, E_{\rm sat,sym}, L_{\rm sym}, K_{\rm sat,sym}, Q_{\rm sat,sym}, Z_{\rm sat,sym} \}$, for a meta-model run at order $N=4$, the Landau nucleon effective mass at nuclear saturation density in symmetric matter, $m_{\rm sat}^\star/m$, and the effective mass isosplit, $\Delta m_{\rm sat}^\star/m$, defined at nuclear saturation as the difference between the neutron and proton Landau effective mass in pure neutron matter. To allow for additional freedom of the functional at high density and to reproduce any arbitrary nuclear model in a large density domain, we considered different values for the third- and fourth-order nuclear empirical parameters below and above saturation density.
We denote these values $(Q_{\rm sat,sym}, Z_{\rm sat,sym})$ below saturation and $(Q^\star_{\rm sat,sym}, Z^\star_{\rm sat,sym})$ above saturation, respectively, as done in \citet{Mondal2023} (see also \citet{Klausner2025b}).
As a consequence, while $Q$ and $Z$ are, as in the customary definition, associated to the (third and fourth) derivatives of the energy functional at saturation, $Q^\star$ and $Z^\star$ have to be rather interpreted as effective parameters (see also the discussion in Sect.~III.E in \citet{Margueron2018}). In this way, we ensure that the low-density behaviour of the equation of state, mainly constrained by observables sensitive to sub-saturation density, does not impose any spurious correlation to the high-density regime. 
Since these parameters are changed for each model at $n_{\rm sat}$, this procedure does not yield discontinuities in the pressure or sound speed. In addition, for the present analysis, the $b$ parameter governing the low-density limit (see Eq.~(\ref{eq:uk})) is varied in the $[1,10]$ range.
Each set of these latter (meta-model) parameters is complemented, for the non-homogeneous crust computed with the CLDM approach described in Sect.~\ref{sec:cldm}, by five surface and curvature parameters.
Specifically, four of them, $(\sigma_0, \sigma_{0,{\rm c}}, b_{\rm s}, \beta)$, were optimised to reproduce the AME2020 experimental mass table, while the surface parameter $p$ governing the isospin behaviour of the surface tension was fixed at $p=3$, as done in \citet{Dinh2021c} and \citet{Davis2024} (see e.g. \citet{Carreau2019, Dinh2021a} for a discussion). The complete set of 17 parameters is thus denoted as $\mathbf{X}$. We considered two prior sets in our Bayesian analysis: 
\begin{enumerate}[(i)]
    \item Each parameter was randomly sampled from a uniform flat distribution whose minimum and maximum values are listed in Table~\ref{tab:nuc-emp-param}. We note that the intervals for the parameters $m^\star_{\rm sat}$ and $\Delta m^\star_{\rm sat}$ are different with respect to those used in previous works based on the same approach (see e.g. \citet{Dinh2021a,Dinh2021b,Dinh2021c,Davis2024, Burrello2025}). The reason for enlarging these ranges was to increase flexibility and have the ability to make meaningful comparisons of our posterior distributions with those obtained in \citet{Klausner2025} (see point ii and Fig.~\ref{fig:gaussian}).
    
    \item The parameters $\{ n_{\rm sat}$, $E_{\rm sat,sym}$, $L_{\rm sym}$, $K_{\rm sat,sym}$, $Q_{\rm sat,sym}$, $Z_{\rm sat,sym}$, $m_{\rm sat}^\star$, $\Delta m_{\rm sat}^\star \}$ were drawn within the same intervals as in Table~\ref{tab:nuc-emp-param} but from a multivariate Gaussian distribution fitting the posterior distributions obtained in \citet{Klausner2025} by considering a comprehensive set of static and dynamic nuclear-structure observables. These include nuclear masses and charge radii, spin-orbit splittings, electric dipole polarizability and parity-violating asymmetry, excitation energy of the isoscalar giant monopole resonance, energy-weighted sum rule of the isovector giant dipole resonance, and excitation energy of the isoscalar quadrupole resonance. The other parameters of the model, that is, $Q^\star_{\rm sat,sym}$, $Z^\star_{\rm sat,sym}$, and $b$, were drawn as described in point (i). Details of this implementation are given in Appendix~\ref{sec:gauss}.
\end{enumerate}

\begin{table}[]
\caption{Minimum and maximum values of the parameter set $\mathbf{X}$.}
\centering
\begin{tabular}{lcc}
\hline
Parameter & min &  max  \\
\hline
$E_{\text {sat }}[\mathrm{MeV}]$ & -17 & -15 \\
$n_{\text {sat }}\left[\mathrm{fm}^{-3}\right]$ & 0.15 & 0.17  \\
$K_{\text {sat }}[\mathrm{MeV}]$ & 190 & 270 \\
$Q_{\text {sat }}[\mathrm{MeV}]$ & -1000 & 1000 \\
$Z_{\text {sat }}[\mathrm{MeV}]$ & -3000 & 3000 \\
$E_{\text {sym }}[\mathrm{MeV}]$ & 26 & 38 \\
$L_{\text {sym }}[\mathrm{MeV}]$ & 10 & 80 \\
$K_{\text {sym }}[\mathrm{MeV}]$ & -400 & 200 \\
$Q_{\text {sym }}[\mathrm{MeV}]$ & -2000 & 2000 \\
$Z_{\text {sym }}[\mathrm{MeV}]$ & -5000 & 5000 \\
$Q_{\text {sat }}^{\star} [\mathrm{MeV}]$ & -2000 & 2000 \\
$Z_{\text {sat }}^{\star} [\mathrm{MeV}]$ & -5000 & 5000 \\
$Q_{\text {sym }}^{\star} [\mathrm{MeV}]$ & -3000 & 3000 \\
$Z_{\text {sym }}^{\star} [\mathrm{MeV}]$ & -5000 & 5000 \\
$m_{\text {sat }}^{\star} / m$ & 0.6 & 1.2 \\
$\Delta m_{\text {sat }}^{\star} / m$ & 0.0 & 2.0 \\
$b$ & 1 & 10 \\
\hline
\end{tabular}
\label{tab:nuc-emp-param}
\end{table}

Each prior model is then passed through different filters, thus the total likelihood of each model (and of each stellar property thus calculated) is given by 
\begin{equation}
    \mathcal{L}(\mathbf{X}) = \prod_j \mathcal{L}_j = \prod_j p(c_j|\textbf{X}) \ ,
    \label{eq:likelihood}
\end{equation}
with $p(c_j|\textbf{X})$ being the conditional probability of reproducing the constraint $c_j$ assuming the parameter set $\mathbf{X}$.
The following filters were applied (see also Appendix~\ref{app:bayes} and \citet{Dinh2021c, Davis2024, Montefusco2025} for details):
\begin{enumerate}
    \item Models have to result in meaningful solutions for the crust. In other words, the minimisation procedure (see Sect.~\ref{sec:cldm}) has to lead to positive neutron-gas and cluster densities.
    
    \item Nuclear masses. For each set of parameters $\mathbf{X}$, the surface parameters in the CLDM were fitted to the AME2020 \citep{ame2020} experimental masses. Parameter sets that yielded a (non-)convergent fit were retained (discarded); thus, the corresponding pass-band filter $\omega_{\rm AME}(\mathbf{X})$ is set to 1 (0). The likelihood of each (successful) model was then quantified by the quality of the reproduction of the nuclear masses, defined from the error estimator $\chi^2(\mathbf{X},\sigma_0,\sigma_{0,{\rm c}},b_s,\beta)$ (see Eqs.~\eqref{eq:filter-masses}-\eqref{eq:chi2-masses}). When using the prior described in point (ii), a given parameter set $\mathbf{X}$ still had to yield a convergent fit to the AME2020, but the $\chi^2$ weight was not applied, since experimental nuclear-mass data were already included among the set of nuclear properties considered in \citet{Klausner2025} and the corresponding likelihood is reflected in the multivariate Gaussian distributions from which the parameter set was drawn.

    \item Low-density ab initio calculations. The nuclear model defined by the parameter set, $\mathbf{X}$, has to be consistent with the energy per nucleon of pure neutron matter obtained by ab initio calculations using chiral effective interactions ($\chi$-EFT) and renormalisation group methods. We used the energy band in \citet{Huth2021} (see their Fig.~1), which represents a conflation of results available in the literature, in the density range $n_B \in [0.02;0.2]$~fm$^{-3}$. We applied a corresponding likelihood that is flat inside the energy band and has a Gaussian tail outside (see Eqs.~\eqref{eq:filter-EFT}, \eqref{eq:filter-ETF-gauss}, and \eqref{eq:filter-ETF-Qn}).
    
    \item Causality and thermodynamic stability. This pass-band filter, effective at high density, ensures that each model satisfies causality, thermodynamical stability, and that the symmetry energy is non-negative at all densities; thus $\mathcal{L}_{\rm HD}(\mathbf{X}) = \omega_{\rm HD}$, with $\omega_{\rm HD}(\mathbf{X}) = 1\ (0)$ if the model is retained (discarded). 

    \item NS maximum mass. This filter accounts for observational data of massive NSs, specifically of the precisely measured masses of PSR J0348$+$0432, $M_{J0348} = 2.01 \pm 0.04$\;M$_{\odot}$ \citep{Antoniadis2013}, PSR J1614$-$2230, $M_{J1614} = 1.908 \pm 0.016$\;M$_{\odot}$ \citep{Arzoumanian2018}, and PSR J0740$+$6620, $M_{J0740} = 2.07 \pm 0.07$\;M$_{\odot}$ \citep{Miller2021}. The associated likelihood is defined as the convolution of cumulative Gaussian distribution functions (see Eq.~\eqref{eq:filter-Mmax}). 

    \item Tidal deformability from GW170817. We included a filter accounting for the tidal deformability data from the GW170817 event obtained by the LIGO-Virgo-KAGRA collaboration \citep{Abbott2019prx}. The associated likelihood is given by Eq.~\eqref{eq:filter-tidal}.
    
    \item NICER data. We implemented constraints from the mass--radius X-ray pulse-profile estimates from the NICER collaboration. Specifically, we considered the following sources: PSR J0030$+$0451 \citep{Miller2019}, PSR J0740$+$6620 \citep{Miller2021}, PSR J0437$-$4715 \citep{Choudhury2024}, and PSR J0614$-$3329 \citep{Mauviard2025}. The associated likelihood is given by Eq.~\eqref{eq:filter-nicer}.
    
\end{enumerate}

\subsection{Uncertainties in the shear modulus and speed}
\label{sec:res-shear}

\begin{figure*}[!ht]
    \sidecaption
    \includegraphics[width=12cm]{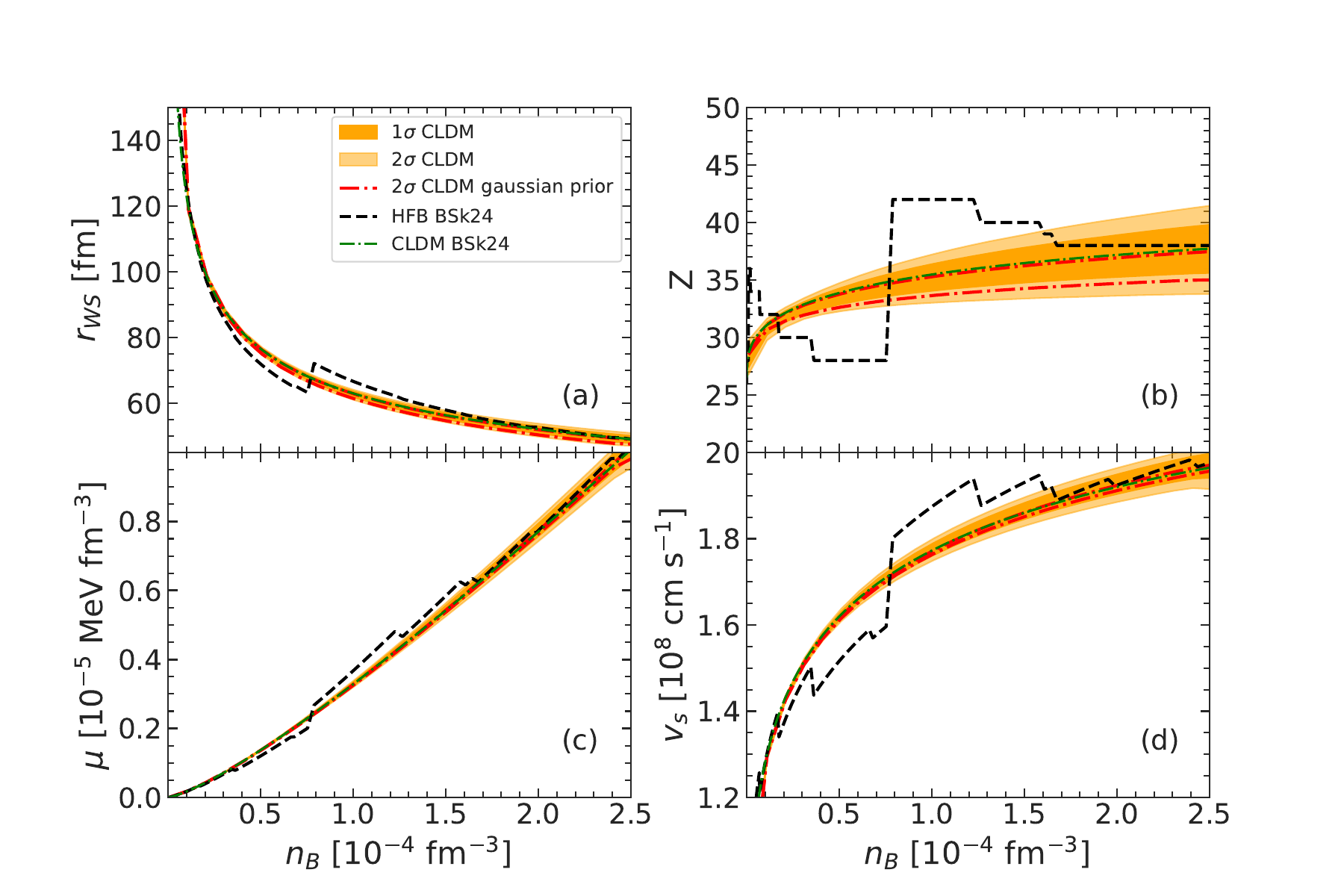}
    \caption{Wigner-Seitz-cell radius (panel a), proton number $Z$ of the ions (panel b), shear modulus $\mu$ (panel c), and shear speed (panel d) as a function of baryon number density in the outer crust. Shaded areas correspond to $1\sigma$ and $2\sigma$ confidence intervals for the uniform prior, while dash-dotted red lines delimit the $2\sigma$ confidence interval for the non-uniform prior. Dashed black (green dash-dotted) lines correspond to results of \citet{Pearson2018} (CLDM model with the BSk24 parameters). See text for details.}
    \label{fig:shear-outer}
\end{figure*}

We start our discussion with a presentation of the elastic properties of the outer crust in Fig.~\ref{fig:shear-outer}. The dark (light) yellow shaded areas represent the $1 \sigma$ ($2 \sigma$) confidence intervals of our posterior when all filters (1. to 7.) are included, starting from a uniform prior for the parameter set $\mathbf{X}$ (point i in Sect.~\ref{sec:bayes}), while the red dash-dotted lines identify the $2\sigma$ confidence interval obtained starting from a non-uniform prior (point ii in Sect.~\ref{sec:bayes}).
For comparison, we also show the CLDM results obtained using the BSk24 empirical parameters (green dash-dotted lines) and the results from the outer-crust data in \citet{Pearson2018}, for which the ion masses have been calculated using experimental mass data whenever available complemented with theoretical masses from the HFB-24 mass model \citep{Goriely2010} (black dashed lines). In all panels, we can observe a similar trend in the results for the shown crust properties; namely, the WS radius $r_{\rm WS}=(3V_{\rm WS}/(4\pi))^{1/3}$ (panel a), the ion proton number $Z$ (panel b), the shear modulus (panel c), and the shear speed (panel d).
The results presented in Fig.~\ref{fig:shear-outer} were obtained by neglecting the correction factor in the shear modulus, that is, using Eq.~\eqref{eq:shear-mod}.  Indeed, we checked that the effect of this correction is negligible in the density range explored by the outer crust, thus the two results would be indistinguishable in the figure. 
The increase in the shear modulus is mainly due to the decrease of $r_{\rm WS}$ with increasing density in the crust, combined with the slight increase in $Z$; see Eq.~\eqref{eq:shear-mod}. As for the shear speed, its increase is mainly due to the increase in $\mu$ (we recall that in the outer crust $\rho_{B,{\rm g}} =0$, thus $\rho_{B,{\rm d}} = \rho_B$).
The CLDM results using the BSk24 empirical parameters are in very good agreement with our posterior at $1\sigma$, as previously noticed for other NS properties in \citet{Dinh2021a} and \citet{Fantina2023}, while variations can be observed with respect to the more microscopic calculations of \citet{Pearson2018} based on the same functional. This reflects the inclusion of nuclear-structure effects in \citet{Pearson2018}, which are absent in the CLDM approach.
The most striking feature observed in Fig.~\ref{fig:shear-outer} is the reduction of the $2 \sigma$ uncertainty band when experimental nuclear data are included in the prior. This can be seen when comparing the red dash-dotted lines to the light-yellow shaded area. In any case, the uncertainties in the NS outer-crust properties are relatively narrow, since the only nuclear inputs in their determination are nuclear masses (either experimentally measured or theoretically calculated) thus the model dependence is relatively small and mainly affects the deeper layers.

\begin{figure*}[!ht]
    \sidecaption
    \includegraphics[width=12cm]{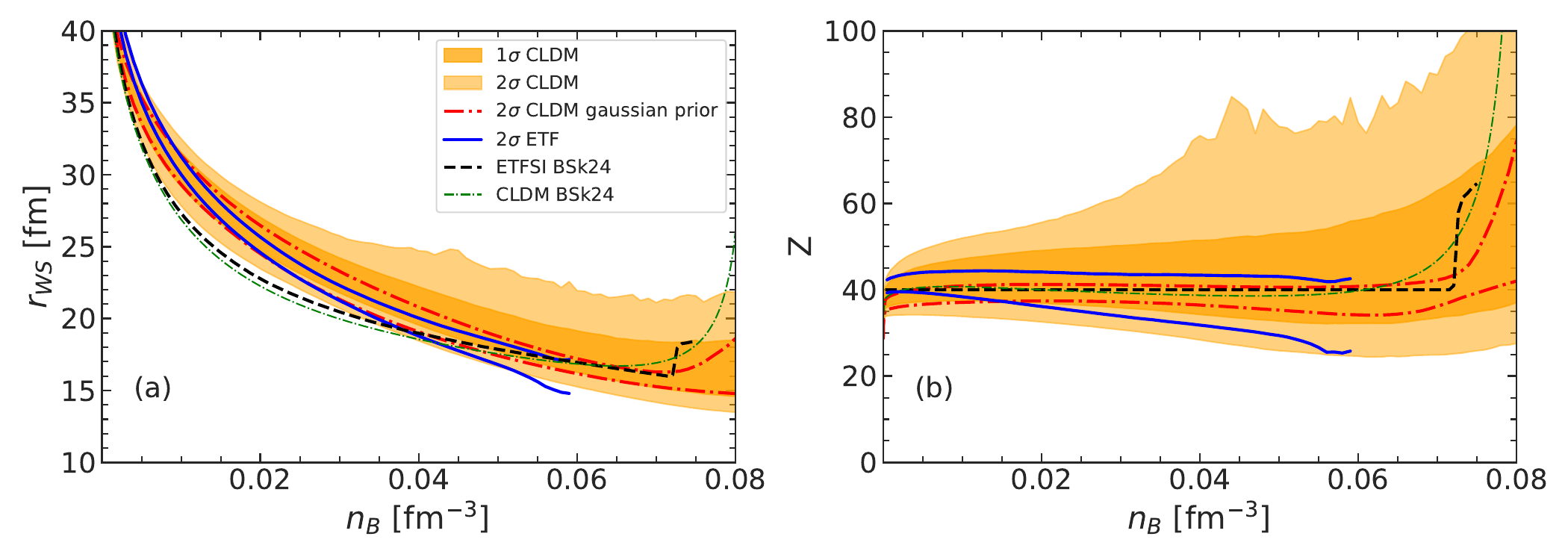}
    \caption{Same as in Fig.~\ref{fig:shear-outer} for the Wigner-Seitz-cell radius (panel a) and proton number $Z$ of the ions (panel b) in the inner crust. Solid blue lines delimit the $2\sigma$ confidence interval of \citet{Klausner2025}. Shear speed is computed from the dynamical mass in Eq.~\eqref{eq:rhobd}. See text for details.}
    \label{fig:inner-comp}
\end{figure*}

\begin{figure*}[!ht]
    \sidecaption
    \includegraphics[width=12cm]{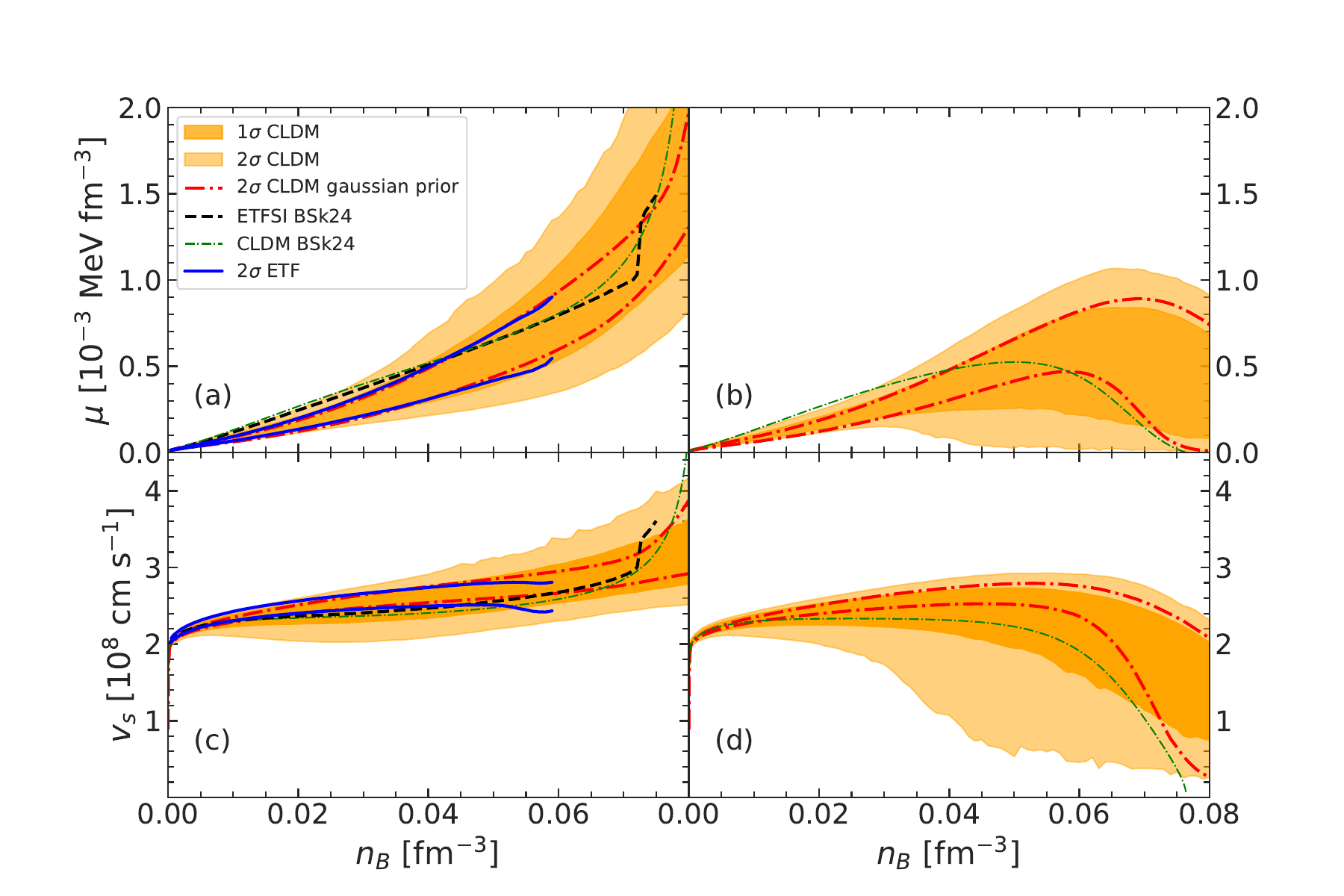}
    \caption{Same as in Fig.~\ref{fig:inner-comp} for the shear modulus (panels a and b) and the shear speed (panels c and d). Results are obtained using Eq.~\eqref{eq:shear-mod} (panels a and c) or Eq.~\eqref{eq:shear-mod-corr} (panels b and d) for the shear modulus.}
    \label{fig:shear-inner}
\end{figure*}

On the other hand, as expected, the uncertainties are larger in the inner crust, as shown in Figs.~\ref{fig:inner-comp} and \ref{fig:shear-inner}. As in Fig.~\ref{fig:shear-outer}, the $1\sigma$ and $2\sigma$ confidence interval obtained with the CLDM (yellow shaded areas) starting from a uniform prior (point i in Sect.~\ref{sec:bayes}) are displayed together with the $2\sigma$ confidence interval obtained starting from a nuclear-physics-informed prior (point ii in Sect.~\ref{sec:bayes}; red dash-dotted lines).
For comparison, we also show the $2 \sigma$ confidence interval resulting from the posterior models of \citet{Klausner2025b} (blue lines)\footnote{In \citet{Klausner2025b}, the implemented constraint on the maximum mass was that of PSR J0348$+$0432 \citep{Antoniadis2013} only, while we used also the observational data from PSR J1614$-$2230 \citep{Arzoumanian2018} and PSR J0740$+$6620 \citep{Miller2021}; see Sect.~\ref{sec:HD-filters}. However, the impact of this choice is expected to be negligible on the crust results presented here.}, where the crust was treated in the extended Thomas-Fermi (ETF) approach, as well as the results obtained using the inner-crust data of \citet{Pearson2018} computed with the ETF plus Strutinski integral (ETFSI) calculations based on the BSk24 functional (black dashed lines).
We note that the blue curves stop at $n_B \approx 0.06$~fm$^{-3}$ because of convergence issues, as explained in \citet{Klausner2025b}.
From Fig.~\ref{fig:inner-comp}, we can again observe in all approaches a similar trend, that is, a decrease of the WS radius with density, while $Z$ remains fairly constant for most of the densities explored in the inner crust, and only increases sensibly at the bottom of the crust towards the crust-core transition (see panels a and b, respectively), in agreement with previous works (see e.g. \citet{Carreau2020, Mondal2020, Grams2022epj}).

As for the elastic properties in the inner crust, from the left panels of Fig.~\ref{fig:shear-inner}, we can note an increase of the shear modulus while the shear speed is only slightly increasing when the expression of the shear modulus for point-like nuclei, Eq.~\eqref{eq:shear-mod}, is used. The behaviour of $\mu$ is mainly due to the decreasing WS radius with density, while the trend seen for $v_s$ can be attributed to the increasing importance of the neutron-gas-density contribution in the dynamical density; see Eq.~\eqref{eq:rhobd}.
Also in the inner crust, the most remarkable effect is the reduction of the $2 \sigma$ uncertainty band when experimental nuclear data are included in the prior (blue solid lines and red dash-dotted lines).
As already observed in \citet{Klausner2025b}, the nuclear-physics-informed prior noticeably decreases the uncertainties in the predictions of the NS crust.
However, we might expect the blue solid curves \citep{Klausner2025b} and the red dash-dotted curves to be almost identical, and for the curves to be superimposed, since the non-uniform prior informed by nuclear-physics observables was employed in obtaining both sets of results.
The use of the multivariate Gaussian distributions as prior for the CLDM calculations (red dash-dotted curves) instead of the original distributions of \citet{Klausner2025} is unlikely to account for the discrepancy since the former distributions reproduce very well those of \citet{Klausner2025}, as discussed in Appendix~\ref{app:bayes} (see Fig.~\ref{fig:gaussian}). The discrepancies between these two sets are thus likely due to the different many-body treatment of the crust, that is, the ETF versus the CLDM approach. Indeed, it was already noticed in \citet{Grams2025} that the use of different finite-size treatments can yield different outcomes for the NS-crust composition and WS-cell radius, which are thus reflected in the shear properties.

However, when Eq.~\eqref{eq:shear-mod-corr} accounting for the finite size of nuclei is employed, the shear modulus, and, thus, the shear speed, are notably reduced in the innermost part of the inner crust, as can be seen from panels (b) and (d) of Fig.~\ref{fig:shear-inner}.
This behaviour can be expected since the configuration with deformed clusters (that gives rise to the correction factor in Eq.~\eqref{eq:shear-mod-corr}) is energetically favoured when the lattice is deformed under shear stress. 
The energy change with respect to the undeformed state is thus lower in the latter case compared to the case of point-like nuclei; thus, the shear modulus is reduced too (see also Figs.~2-3 in \citet{Zemlyakov2023a} and relative discussion). 
We did not show the ETF(SI) results in the right panels of Fig.~\ref{fig:shear-inner} because this would require a definition of the nuclear radius to determine the value of $u$.
Protocols exist to extract these quantities from ETF calculations for comparisons with the CLDM (see e.g. the discussion in \citet{Grams2025}).
However, similarly to panels (a) and (c), we expect that ETF(SI) results inferred from crustal properties in \citet{Klausner2025b} \citep{Pearson2018} closely follow our CLDM predictions obtained with a non-uniform prior (the BSk24 functional).
The median and $1\sigma$ confidence intervals for the shear modulus at different representative densities in the outer and inner crust are reported in Appendix~\ref{app:median} for completeness.

We recall here that we are only considering spherical clusters in this work. The presence of pasta phase is expected to lower the shear modulus and thus the shear speed (see e.g. \citet{Pethick1998, Zemlyakov2023b}).
Therefore, the values that we obtain in the densest regions of the inner crust near the crust--core transition under the hypothesis of spherical cluster without any finite-size correction (left panels in Fig.~\ref{fig:shear-inner}) would constitute an upper limit.
If we included pasta phases\footnote{Pasta phases are actually quite robustly predicted in our formalism, but their appearance was shown to only slightly affect the crust--core transition density \citep{Dinh2021a, Dinh2021b}.} and limited ourselves to the direction-averaged shear modulus (i.e. neglecting transversal shear), taking the commonly employed approximation of a linear interpolation of the shear modulus from its value at the sphere-pasta transition to zero at the crust-core transition (as done e.g. in \citet{Parmar2022}), we would expect to find similar results to those shown in the right panels of Fig.~\ref{fig:shear-inner}. To have a more realistic picture, we would need to extend our calculations to non-isotropic deformations in a future work.

\subsection{Estimation of the frequency of the fundamental mode}
\label{sec:res-freq}

Knowledge of the shear speed also allows us to estimate the frequency of the fundamental $n=0$ oscillation crust mode (see Eq.~\eqref{eq:omega}).
In fact, QPOs discovered in the afterglow following magnetar giant flares, whose frequencies span the Hz to kHz range, have been considered to be associated with NS oscillations.
In particular, torsional shear modes, that have a lower excitation energy compared to other vibrational modes, have been invoked to explain observed QPOs \citep{Duncan1998}. 
We restrict ourselves to the discussion of the fundamental lower order ($l=2$) mode and we compare the estimated frequency with the lowest observed QPO frequencies of SGR 1806$-$20 and SGR 1900$+$14 \citep{Israel2005,Strohmayer2005,Strohmayer2006,Watts2006ApJ}.
Additional NS properties, such as the equation of state and mass-radius relation, computed with a nuclear-physics-informed prior (see point ii in Sect.~\ref{sec:bayes}) are shown and compared with those obtained with a uniform prior (point i in Sect.~\ref{sec:bayes}; see also \citet{Dinh2021c}) in \citet{Klausner2025b}. More details are given in their Figs.~9-12.

To calculate the frequency, Eq.~\eqref{eq:omega}, we considered an average value of the shear speed. Indeed, assigning the value of $v_s$ at a specific density point in the crust, such as the crust--core transition, would yield an overestimation (underestimation) of its value (see lower panels of Fig.~\ref{fig:shear-inner}) and would increase the uncertainties in its determination, as the crust--core transition is known to be model-dependent.
Thus, we calculated the average of the shear speed as
\begin{equation}
    \langle v_s \rangle_{n_B} = \frac{\int_0^{n_{B, {\rm max}}} v_s(n_B) dn_B}{n_{B,{\rm max}}} \ ,
    \label{eq:vs-aver1}    
\end{equation}
where $n_{B,{\rm max}}$ is either the crust-core transition density if the shear modulus is computed with Eq.~\eqref{eq:shear-mod} or the density at which the shear modulus becomes zero if the correction factor is accounted for, as expressed in Eq.~\eqref{eq:shear-mod-corr}. We checked for several nuclear functionals that the discrepancy is less than about $10\%$ between $\langle v_s \rangle_{n_B}$ and the density-weighted average,
\begin{equation}
    \langle v_s \rangle_r = \frac{\int_{R_{\rm min}}^R v_s(r) \rho(r) r^2 dr}{\int_{R_{\rm min}}^R \rho(r) r^2 dr} \ ,
    \label{eq:vs-aver2}    
\end{equation}
with $R_{\rm min}$ being the radius corresponding to $n_{B,{\rm max}}$, $R$ the star radius, and $\rho$ the mass-energy density. 
The reason for using Eq.~\eqref{eq:vs-aver1} instead of Eq.~\eqref{eq:vs-aver2} to estimate the average shear speed mainly lies in the advantage of avoiding the calculation of the integral for each density profile (i.e. for each NS mass) and for each model in the Bayesian analysis, thereby reducing the computational time. 
Moreover, we verified that the variation of the density-weighted shear speed weakly depends on the NS mass (this is also confirmed by other studies, see e.g. Fig.~5 in \citet{Neill2021}) and that the way the average is performed (either via Eq.~\eqref{eq:vs-aver1} or \eqref{eq:vs-aver2}) negligibly affects the estimated frequencies (see Table~\ref{tab:frequencies}). Therefore, we find that the use of a NS-mass-independent value of $\langle v_s \rangle_{n_B}$ (as found using Eq.~\eqref{eq:vs-aver1}) remains a good approximation.

\begin{figure*}[!ht]
    \sidecaption
    \includegraphics[width=12cm]{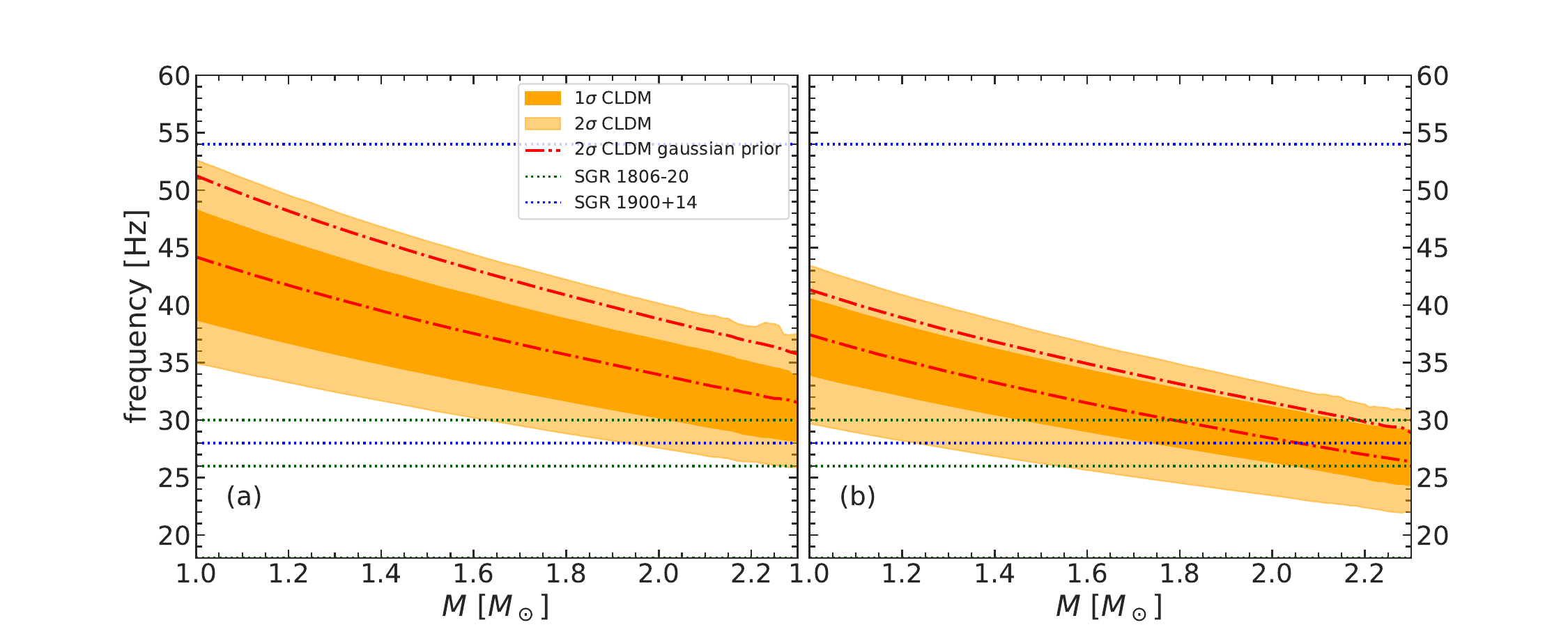}
    \caption{Frequency of the fundamental $_2t_0$ mode as a function of the NS mass obtained using Eq.~\eqref{eq:shear-mod} (panel a) or Eq.~\eqref{eq:shear-mod-corr} (panel b) for the shear modulus. Shaded areas correspond to $1\sigma$ and $2\sigma$ confidence intervals for the uniform prior, while dash-dotted red lines delimit the $2\sigma$ confidence interval for the nuclear-physics-informed prior. Horizontal dotted lines correspond to observed QPO frequencies. See text for details.}
    \label{fig:freq}
\end{figure*}

In Fig.~\ref{fig:freq}, we show the $1\sigma$ and $2\sigma$ confidence interval for the frequency (that is, $\omega_0/(2\pi)$; see Eq.~\eqref{eq:omega}) of the $_2t_0$ mode, as a function of the NS mass. In panel a (panel b) we plot the results obtained using Eq.~\eqref{eq:shear-mod} (Eq.~\eqref{eq:shear-mod-corr}) for the shear modulus.
We display the posterior of our Bayesian analysis using both the uniform (yellow shaded areas) and the non-uniform nuclear-physics-informed prior (red dash-dotted lines).
We do not show the results from more microscopic calculations since, in view of the outcomes in Figs.~\ref{fig:shear-outer} and \ref{fig:inner-comp}, we expect them to lie within our posterior. Horizontal lines corresponds to a chosen set of observed QPO frequencies: 28 and 54~Hz from SGR 1900$+$14 (dotted blue lines) and 18, 26, and 30~Hz from SGR 1806$-$20 (dotted green lines); higher frequencies have been detected but not displayed, as they are likely associated to higher order modes. Indeed, a correct identification of detected QPO frequencies in terms of different oscillation modes remains complicated.
We observe from panel (a) that our posterior would be compatible with frequencies in the range $\approx 25$~Hz to $\approx 50$~Hz ($\approx 30$~Hz to $\approx 50$~Hz in the case of non-uniform prior). This could possibly disfavour the shear mode for the lowest frequency, at 18~Hz.
The overall reduction of the frequency seen in panel (b) is a direct consequence of the reduction of $v_s$ when the correction to the shear modulus is implemented (see Eq.~\eqref{eq:shear-mod-corr} and Fig.~\ref{fig:shear-inner}), the posterior predicting in this case frequencies in the range $\approx 20$~Hz to $\approx 40$~Hz ($\approx 25$~Hz to $\approx 40$~Hz for a non-uniform prior).
The median and $1\sigma$ confidence intervals for the frequency for the representative case of a $1.4\ M_\odot$ NS are reported in Appendix~\ref{app:median} for completeness.
We note, however, that Eq.~\eqref{eq:omega} is only an order-of-magnitude estimate; thus, the actual figures have to be taken with care and no incompatibility with observed low frequency ($\lesssim 20$~Hz) can be firmly inferred from our results.

Compared with results in the literature, some earlier studies report torsional frequencies smaller than those shown in Fig.~\ref{fig:freq}. However, a number of previous works calculated the frequency considering full entrainment (i.e. $\rho_{B,{\rm d}} = \rho_B$ in Eq.~\eqref{eq:rhobd}).
This yields an underestimation of the frequency with respect to the case where $\rho_{B,{\rm d}} = \rho_B - \rho_{\rm g}$, that is, where entrainment is taken into account as estimated in recent works \citep{Almirante2025}. For example, using the SLy4 empirical parameters, we obtain a frequency of about $18$~Hz for a $1.4 M_\odot$ if we employ an averaged shear speed (either $\langle v_s \rangle_{r}$ or $\langle v_s \rangle_{n_B}$), in the case of full entrainment and without finite-size corrections (see Table~\ref{tab:frequencies}).
This is in good agreement with the results shown in Fig.~3 of \citet{Steiner2009}, as well as with those in \citet{Tews2017} (see their Fig.~8) and with the recent results reported in \citet{Parmar2026} (see their Fig.~5 for the case without pasta). Indeed, including uncertainties on the composition, the surface parameters, and the entrainment, \citet{Tews2017} reported frequencies for the fundamental $l=2$ mode in the range $\approx 14-52$~Hz, while the Bayesian study of \citet{Parmar2026} indicates frequencies of $\approx 10-23$~Hz for the same mode.
As can be seen in Fig.~8 of \citet{Tews2017}, the main factor contributing to increasing the uncertainties and the overall values of frequencies appears to be the treatment of the entrainment. 

More recently, \citet{Sotanietal2024PRD} presented computations of torsional, $_lt_n$, and shear modes, as well as couplings between axial and polar oscillation modes solving axisymmetric perturbation equations.
Employing the SLy4 interaction and considering a $1.4 M_\odot$, he predicted the frequency of the $_2t_0$ mode at $23.4$~Hz; see their Table~I.
For the same mode, using Eq.~\eqref{eq:omega} and the SLy4 empirical parameters, we obtain a similar value, 30~Hz, when $v_s$ is calculated at the crust-core transition in the case of full entrainment and no finite-size corrections, which should better correspond to the conditions of the calculation by \citet{Sotanietal2024PRD} (see Table~\ref{tab:frequencies}).
On the other hand, our estimate when we include a more realistic treatment of entrainment is noticeably larger, due to the larger value of $v_s$ because of the lower dynamical density.
In both cases, the values are sensibly reduced when the frequencies are estimated using an average value of the shear speed with respect to taking $v_s$ at the crust--core transition, as expected from Fig.~\ref{fig:shear-inner}.
It is interesting to see that overall the different approximations considered (i.e. evaluation of $v_s$ at the crust--core transition, full entrainment, and neglecting the finite-size corrections in the calculation of $\mu$) seem to compensate each other and the resulting frequency, $30$~Hz, is very close to our fiducial value, $33.2$~Hz, calculated with a density-weighted shear speed, Eq.~\eqref{eq:vs-aver2}, using what we think to be the more realistic expression for the entrainment, Eq.~\eqref{eq:rhobd}, and employing Eq.~\eqref{eq:shear-mod-corr} for $\mu$. These values are also compatible with that obtained in the more sophisticated approach of \citet{Sotanietal2024PRD}, $23.9$~Hz (see the last line in Table~\ref{tab:frequencies}).

To check the impact of the different nuclear matter parameters on the frequency of the fundamental modes, we also calculated the Pearson correlation coefficients. In particular, we observe an anti-correlation of the frequency with the slope of the symmetry energy $L_{\rm sym}$, as already reported in previous works (see e.g. \citet{Gearheart2011}, \citet{Sotani2024}, and \citet{Parmar2026}).

\begin{table}[]
\centering
\caption{Frequencies (in Hz) of the $_2t_0$ mode for a $M=1.41 M_{\odot}$ NS under different hypothesis. 
}
\begin{tabular}{cccc}
\hline
 & Correction & Full ent. & No ent. \\
\hline
$v_{s,{\rm cc}}$ & no & 30.0 & 72.7 \\
$\langle v_s \rangle _{n_B}$ & no & 17.4 & 42.1 \\
$\langle v_s \rangle _{n_B}$ & yes & 14.5 & 35.0 \\
$\langle v_s \rangle _{r}$ & no & 17.9 & 43.8 \\
$\langle v_s \rangle _{r}$ & yes & 13.5 & 33.2 \\
\citet{Sotanietal2024PRD} & no & 23.9 & \\ 
\hline
\end{tabular}
\tablefoot{Results are obtained employing the SLy4 empirical parameters using different estimations for the shear velocity: $v_s$ calculated at the crust--core transition (labelled $v_{s,{\rm cc}}$), average value $\langle v_s \rangle _{n_B}$ calculated as in Eq.~\eqref{eq:vs-aver1}, and density-averaged value $\langle v_s \rangle _{r}$ calculated as in Eq.~\eqref{eq:vs-aver2}. Computations either include (labelled `yes') or neglect (labelled `no') the finite-size correction in Eq.~\eqref{eq:shear-mod-corr}, and assume either full entrainment (`Full ent.', that is, $\rho_{B,{\rm d}}=\rho_B$ in Eq.~\eqref{eq:shear-speed}) or entrainment according to \citet{Almirante2025} (`No ent.', that is, $\rho_{B,{\rm d}}=\rho_B - \rho_{B,{\rm g}}$ in Eq.~\eqref{eq:shear-speed}). The last line reports the result of \cite{Sotanietal2024PRD} (see their Table~I).}
\label{tab:frequencies}
\end{table}

\section{Conclusions}
\label{sec:conclusions}

In this work, we performed a Bayesian analysis of shear properties in the NS crust within a meta-model approach for homogeneous nuclear matter complemented with a CLDM for the ions. In agreement with previous studies, we found that the properties of the outer crust are relatively well constrained also for elastic shear properties, while uncertainties increase with density in the inner crust.
In addition, we show that employing a nuclear-physics-informed prior considerably reduces the spread in the prediction for the crust, thus highlighting the importance of nuclear-physics data for the physics of the crust.
Interestingly, the use of a more sophisticated ETF treatment of the WS cell as proposed in \citet{Klausner2025b} leads to results that are very close to those obtained with a simpler CLDM.

The frequencies we estimated for the lower order (torsional) mode, $_2t_0$, are in the range $\sim 25 - 50$~Hz for a uniform prior ($\sim 30 - 50$~Hz in the case of nuclear-physics-informed prior) when the shear modulus is calculated for point-like nuclei, and $\sim 20 - 40$~Hz ($\sim 25 - 40$~Hz) when finite-size corrections in the shear modulus are included. These results remain compatible with observed low-frequency QPOs.
We also note that there are several effects on the shear modulus and speed that are not addressed here, such as electron-screening effects as well as the possible presence of pasta phases (e.g. \citet{Chugunov2022, Zemlyakov2023b, Parmar2022}) and of frozen impurities (e.g. \citet{Fantina2020, Dinh2023aamcp, Kozhberov2019, Potekhin2021, Chugunov2022}) that could alter the value of the estimated frequency and might be included in future works.

\begin{acknowledgements}
This work has been partially supported by the IN2P3 Master Project MAC, and the CNRS International Research Project (IRP) `Origine des \'el\'ements lourds dans l’univers: Astres Compacts et Nucl\'eosynth\`ese (ACNu)'.
The authors would like to thank P. Klausner for providing the data of his publications to generate Figs.~\ref{fig:shear-inner} and \ref{fig:gaussian}, and G. Almirante, M. Antonelli, N. Chamel, H. Dinh Thi, and M. Urban for fruitful discussions.
\end{acknowledgements}

\bibliographystyle{aa}
\bibliography{biblio}

\begin{appendix}

\section{Constraints implemented in the Bayesian analysis}
\label{app:bayes}

We describe additional details on the constraints implemented in the Bayesian analysis presented in Sect.~\ref{sec:bayes}, both coming from nuclear physics (see Sects.~\ref{sec:LD-filters} and \ref{sec:gauss}) and from astrophysics (see Sect.~\ref{sec:HD-filters}); see also \citet{Dinh2021c, Davis2024, Montefusco2025} for details.

\subsection{Low-density filters from nuclear physics}
\label{sec:LD-filters}

In the lower density region, the following filters coming from experimental nuclear physics and theoretical ab initio calculations were applied: 
\begin{enumerate}

    \item \textit{Nuclear masses}. For each set of parameters $\mathbf{X}$, the surface parameters in the CLDM were fitted to the AME2020 \citep{ame2020} experimental masses. Parameter sets for which the fit was successful were assigned a likelihood, quantified by the quality of the reproduction of the nuclear masses, 
    \begin{equation}
        \mathcal{L}_{\rm AME}(\mathbf{X}) = \omega_{\rm AME} \ e^{-\chi^2(\mathbf{X})/2} \ ,
        \label{eq:filter-masses}
    \end{equation}
    with 
    \begin{equation}
        \chi^2(\mathbf{X}) = \frac{1}{N_{\rm dof}} \sum_{j=1}^{N_{\rm AME}} \frac{\left( M^{(j)}_{\rm theo,CLD}(\mathbf{X}) - M^{(j)}_{\rm AME} \right)^2}{\sigma_{\rm th}^2} \ ,
        \label{eq:chi2-masses}
    \end{equation}
    where the sum runs over the nuclei in the AME2020 mass table with $N,Z \ge 8$ ($N=A-Z$), $M_{\rm theo,CLD}(\mathbf{X})$ is the theoretical mass calculated with the CLDM using the parameter set $\mathbf{X}$ complemented with the best-fit surface parameters, 
	\begin{eqnarray}
    M_{\rm theo, CLD} c^2 &=& m_p c^2 Z + m_n c^2 (A-Z) \nonumber \\ 
    &+& A e_{\rm nuc} +4\pi r_N^2\left (\sigma_{\rm s} +\frac{2\sigma_{\rm c}}{r_N}\right ) \nonumber \\ 
    &+& \frac{3}{5}\frac{e^2 Z^2}{r_N} \ ,
    \label{eq:mass-theo}
    \end{eqnarray} 
    $M_{\rm AME}$ is the experimental mass, $\sigma_{\rm th} = 0.04$~MeV/$c^2$ is a systematical theoretical error, and $N_{\rm dof} = N_{\rm AME} - 4$ is the number of degrees of freedom.     
    When starting from a non-uniform prior (point ii in Sect.~\ref{sec:bayes}), the fit still has to yield a convergent solution ($\omega_{\rm AME}=1$), but the weight in Eq.~\eqref{eq:chi2-masses} was not applied since experimental mass measurements were already included in the analysis by \citet{Klausner2025} and the corresponding likelihood is reflected in the posterior distribution from which the parameters were drawn; see Sect.~\ref{sec:gauss}.

    \item \textit{Low-density ab initio calculations}. We considered for this filter the energy per nucleon of pure neutron matter obtained by ab initio calculations using chiral effective interactions ($\chi$-EFT) and renormalisation group methods in \citet{Huth2021} (see their Fig.~1; see also Fig.~4 in \citet{Huth2022}). The band, conflating different results available in the literature, was applied in the density range $n_B \in [0.02;0.2]$~fm$^{-3}$. 
    We divided the density range in $N_n = 10$ equally spaced slices. The associated probability is given by
    \begin{equation}
        \mathcal{L}_{\chi {\rm EFT}} \propto \prod_{j=1}^{N_n} p \left(e_{\mathbf{X}}(n_j)|n_j \right) \ ,
        \label{eq:filter-EFT}
    \end{equation}
    where $e_{\mathbf{X}}(n_j)$ is the pure neutron-matter energy computed with the meta-model characterised by the parameter set $\mathbf{X}$, and we used the Gaussian-like normalised probability distribution
    \small
    \begin{equation}
        p(e|n)= Q_n 
        \begin{cases} \text{exp}\left(-\frac{(e(n)-e_-(n))^2}{2\sigma_n^2}\right) & \text { if } e(n) < e_-(n) \\ 1 & \text { if } e_-(n) \le e(n) \le e_+(n) \\ \text{exp}\left(-\frac{(e(n)-e_+(n))^2}{2\sigma_n^2}\right) & \text { if } e(n)>e_+(n)
        \end{cases}
        \label{eq:filter-ETF-gauss}
    \end{equation}
    \normalsize
    with 
    \begin{eqnarray}
        Q_n &=& \frac{\Delta}{e_+(n)-e_-(n)} \ , \label{eq:filter-ETF-Qn} \\ \sigma_n &=& \frac{1- \Delta}{\Delta \sqrt{2\pi}}[e_+(n)-e_-(n)] \ ,
         \label{eq:filter-ETF-sigma}
    \end{eqnarray}
    $e_-(n)$ ($e_+(n)$) denoting the lower (upper) limit of the conflated energy-per-nucleon band at a given density point $n_j$.
    We set $\Delta=0.9$, meaning that the central plateau of the distribution (the two tails) accounts for the $90\%$ (the remaining $10\%$), in accordance with other prescriptions used in previous studies (\citet{Dinh2021c, Davis2024}; see also Appendix~A in \citet{Montefusco2025} for details). We also normalised the probability $p(e|n)$ within the energy band to 1.
    We checked that this procedure yields results comparable with enlarging the energy band of $5\%$, in agreement with the results in \citet{Scurto2024}.

\end{enumerate}

\subsection{Nuclear-physics informed prior including experimental data: multivariate Gaussian}
\label{sec:gauss}

In the work of \citet{Klausner2025}, a Bayesian inference analysis was performed using a large ensemble of Skyrme functionals. A broad set of static and dynamic nuclear experimental observables were used as constraints. These latter include nuclear masses and charge radii, spin-orbit splittings, excitation energies of the isoscalar giant monopole and quadrupole resonances, the energy-weighted sum rule of the isovector giant dipole resonance, as well as the electric dipole polarizabilities of $^{208}$Pb and $^{48}$Ca.
Additionally, parity-violating asymmetries $A_\text{PV}$ measured in the PREX-II and CREX experiments were incorporated (see \citet{Klausner2025} for details and references therein for the employed constraints). 

\begin{figure*}[!ht]
    \centering
    \includegraphics[scale=0.7]{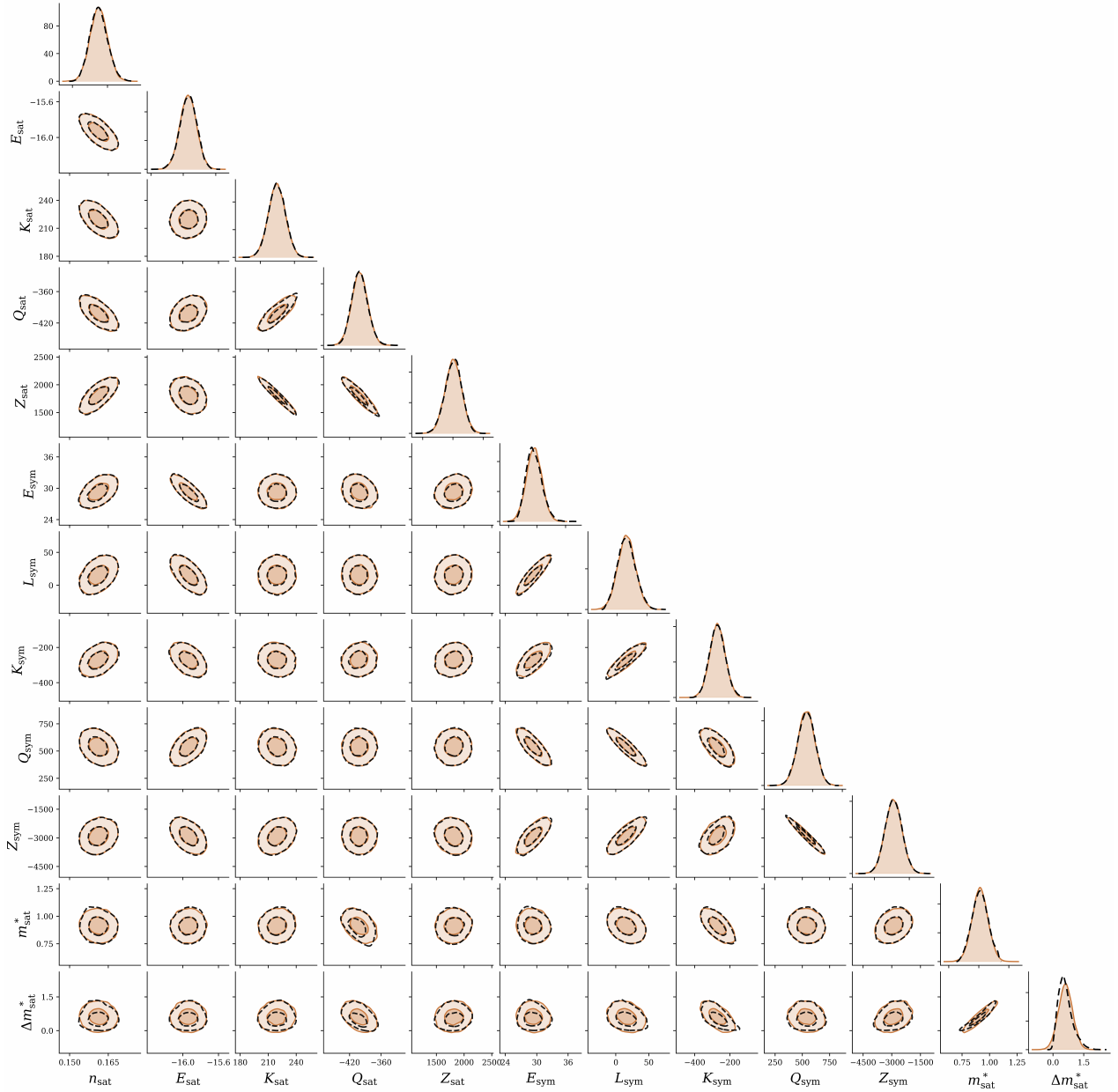}
    \caption{Comparison between the multivariate Gaussian distributions (orange filled areas) and the posteriors from \citet{Klausner2025} (see also \citet{Klausner2025b}; black dashed lines).}
    \label{fig:gaussian}
\end{figure*}

From the resulting posterior distribution over the Skyrme parameters, nuclear empirical parameters can be extracted via analytical mapping. 
This mapping allows to use the posteriors obtained in \citet{Klausner2025} as nuclear-physics-informed priors (or likelihood) for subsequent Bayesian analyses. 
A particularly effective method to incorporate this information is to approximate the posterior distribution of the nuclear empirical parameters using a multivariate Gaussian distribution. 
This is achieved by fitting the mean vector of the nuclear empirical parameters $\mathbf{X}$, $\langle \mathbf{X} \rangle$, and the covariance matrix $\mathcal{M}$ of the posterior, thereby constructing the multivariate Gaussian likelihood, 
\begin{equation}
\small
\mathcal{L}_{\rm nuc}(\mathbf{X}) = \frac{1}{(2\pi)^{n/2} |\mathcal{M}|^{1/2}} \exp\left( -\frac{1}{2} (\mathbf{X} - \langle \mathbf{X} \rangle)^\top \mathcal{M}^{-1} (\mathbf{X} - \langle \mathbf{X} \rangle) \right) \ ,
\normalsize
\end{equation}
where $n$ is the number of parameters. This approximation provides a compact and fully analytical representation of the posterior, preserving both the marginal distributions and the correlations between parameters, without requiring computationally expensive interpolation or kernel density estimation. 
As shown in Fig.~\ref{fig:gaussian}, the multivariate Gaussian approximation (orange filled areas) closely reproduces the original posterior from \citet{Klausner2025} (black dashed lines), and the relative correlations, for almost all parameters.
Some deviations occur for the effective mass, $m^\star_{\rm sat}$, and the effective mass splitting, $\Delta m^\star_{\rm sat}$.
Particularly for the latter quantity, this discrepancy can be attributed to the fact that the original distribution is skewed due to the constraint $\Delta m^\star_{\rm sat} > 0$. This latter constraint cannot be captured by a symmetric Gaussian distribution, leading to a reduced accuracy in the low-value region of $\Delta m^\star_{\rm sat}$.

\subsection{High-density filters from astrophysics}
\label{sec:HD-filters}

In the higher density region, the following filters coming from astrophysical observations were implemented: 
\begin{enumerate}
    
    \item \textit{NS maximum mass}. This filter accounts for observational data of the precisely measured masses of PSR J0348$+$0432, $M_{J0348} = 2.01 \pm 0.04$\;M$_{\odot}$, with $\sigma_{J0348}=0.04$ being the standard deviation \citep{Antoniadis2013}, PSR J1614$-$2230, $M_{J1614} = 1.908 \pm 0.016$\;M$_{\odot}$, with $\sigma_{J1614}=0.016$ \citep{Arzoumanian2018}, and PSR J0740$+$6620, $M_{J0740} = 2.07 \pm 0.07$\;M$_{\odot}$, with $\sigma_{J0740}=0.07$ \citep{Miller2021}. The associated likelihood is defined as the convolution of cumulative Gaussian distribution functions,
    \begin{equation}
        \mathcal{L}_{M_{\rm max}} = \prod_{J} \frac{1}{\sigma_J \sqrt{2 \pi}} \int_0^{\frac{M_{\rm max}(\mathbf{X})}{M_\odot}} e^{-\frac{(x-M_{J}/M_\odot)^2}{2 \sigma_{J}^2}} dx \ ,
        \label{eq:filter-Mmax}
    \end{equation}
     where $J$ runs over the three considered mass measurements, $M_{\rm max}(\mathbf{X})$ is the maximum NS mass at equilibrium (compatible with causality), determined for each model from the solution of the Tolman-Oppenheimer-Volkoff (TOV) equations\footnote{At variance with \citet{Montefusco2025}, the maximum NS mass at equilibrium calculated here and compatible with causality does not necessarily coincide with the maximum TOV mass. Indeed, as in other studies employing (semi-)agnostic models (see e.g. Figs.~1-3 in \citet{Margueron2018b}), violation of causality can occur before reaching the maximum TOV mass for some parameter sets.} \citep{Tolman1939, Oppenheimer1939}, and $M_J$ and $\sigma_J$ are the mass and standard deviation of each mass measurement, respectively.

    \item \textit{Tidal deformability from GW170817}. The likelihood associated to the filter accounting for tidal deformability data from GW170817 obtained by the LIGO-Virgo-KAGRA collaboration is written as
    \begin{equation}
        \mathcal{L}_{\rm GW} = \sum_i P_{\rm GW}(\tilde{\Lambda}(q^{(i)}),q^{(i)}) \ ,
        \label{eq:filter-tidal}
    \end{equation}
    where $q$ is the mass ratio of the lighter object over the heavier one and $P_{\rm GW}(\tilde{\Lambda}(q^{(i)}),q^{(i)})$ is the joint posterior distribution reported in  \citet{Abbott2019prx}\footnote{Data of the posterior distributions are taken from the LIGO Document P1800061-v11 `Properties of the binary neutron star merger GW170817' at \url{https://dcc.ligo.org/LIGO-P1800061/public}, obtained using the PhenomPNRT waveform, referred to as `reference model' in \citet{Abbott2019prx}, and low-spin prior.}.
    In this work, as it was done in \citet{Dinh2021c}, $q$ was chosen to be in the one-sided $90\%$ confidence interval obtained in \citet{Abbott2019prx}, $q \in [0.73, 1.00]$ (see Appendix~B in \citet{Montefusco2025} for details). 
    
    \item \textit{NICER data}. We considered data from X-ray pulse-profile of pulsars PSR J0030$+$0451, PSR J0740$+$6620, PSR J0437$-$4715, and PSR J0614$-$3329 from NICER. The NICER likelihood probability is given by
    \begin{eqnarray}
        \mathcal{L}_{\rm NICER} = \prod_{J} \sum_i P_{J}(M_{J}^{(i)},R_{J}^{(i)}) \ ,
        \label{eq:filter-nicer}
    \end{eqnarray}
    where $P_J(M,R)$ is the joint probability distribution of mass and radius for the pulsar $J$, namely PSR J0030$+$0451 (PSR J0740$+$6620), obtained using NICER (NICER and XMM-Newton) data and the waveform model with three uniform oval spots by \citet{Miller2019} (\citet{Miller2021}), PSR J0437$-$4715 \citep{Choudhury2024}, and PSR J0614$-$3329 \citep{Mauviard2025}. The intervals of ${M_{J0030} = [1.0, 2.2]} M_\odot$ and ${M_{J0740} = [1.68,2.39]} M_\odot$ were chosen to be sufficiently large so that they cover the associated joint mass-radius distributions (see also Fig.~7 in \citet{Miller2019} and Fig.~1 in \citet{Miller2021}). More recently, \citet{Salmi2022} \citep{Vinciguerra2024} performed an analysis of PSR J0740$+$6620 (J0030$+$0451) from NICER data, leading to results ---in particular the inferred mass and radius--- consistent with previous findings and having similar uncertainties. Therefore, we used the constraints from \citet{Miller2021}, as was done in \citet{Dinh2021c} and \citet{Davis2024}. For the PSR J0437$-$4715 and PSR J0614$-$3329, the full kernel density estimation (KDE) was used. 
   
\end{enumerate}

\section{Median values of the shear properties}
\label{app:median}

In this appendix, we report the median and $1\sigma$ confidence levels for the shear properties calculated in this work. In Table~\ref{tab:median}, we list the values obtained for the shear modulus at three chosen densities, namely $10^{-4}$~fm$^{-3}$, $10^{-2}$~fm$^{-3}$, and $0.05$~fm$^{-3}$, that correspond to typical densities encountered, respectively, at the bottom of the outer crust, in the inner crust, and at the bottom of the inner crust; see Figs.~\ref{fig:shear-outer} and \ref{fig:shear-inner}.
The value of $0.05$~fm$^{-3}$ also corresponds to the density where pasta phases are expected to appear.
Indeed, although we considered only spherical clusters in this work, the results for the shear modulus at this density can give an estimate of the uncertainties in the shear modulus at the sphere--pasta transition, as it would be predicted in our model (see e.g. Fig.~4 in \citet{Dinh2021a} and Fig.~3 in \citet{Dinh2021b}).
We can see that the uncertainties in the shear modulus increase with density, and are significantly reduced when the nuclear-physics-informed prior is employed, as already noticed in Figs.~\ref{fig:shear-outer} and \ref{fig:shear-inner}.

We also report in the last row of Table~\ref{tab:median} the median and $1\sigma$ confidence interval for the estimated frequency of the $_2t_0$ mode for a $M=1.4 M_{\odot}$ NS; see Fig.~\ref{fig:freq}. In this case as well, the use of the nuclear-physics-informed prior instead of the uniform prior yields smaller uncertainties, in accordance with the behaviour already noticed for the shear modulus.
We note that although the range of frequencies obtained in this work is in relatively good agreement with observed QPOs frequencies shown in Fig.~\ref{fig:freq}, we think that comparison with data should be handled with care since values for the frequencies obtained via Eq.~\eqref{eq:omega} are only order-of-magnitude estimates.

\begin{table}[!h]
\centering
\caption{Median and $1 \sigma$ confidence intervals of the shear modulus (in MeV~fm$^{-3}$) at different densities and of the frequency (in Hz) of the $_2t_0$ mode for a $M=1.4 M_{\odot}$ NS.
}
\begin{tabular}{ccc}
\hline
 & w/o finite size & with finite size  \\
\hline
\multirow{2}{*}{$\mu (n_B=10^{-4})$} & $3.292^{+0.052}_{-0.060} \times 10^{-6}$ &  $3.292^{+0.052}_{-0.060} \times 10^{-6}$ \\
                                    & $3.2621^{+0.0074}_{-0.0076} \times 10^{-6}$ & $3.2621^{+0.0074}_{-0.0076} \times 10^{-6}$ \\
\hline
\multirow{2}{*}{$\mu (n_B=10^{-2})$} & $8.5^{+2.0}_{-1.6} \times 10^{-5}$ &  $8.5^{+2.0}_{-1.6} \times 10^{-5}$ \\
                                   & $7.49^{+0.82}_{-0.86} \times 10^{-5}$ & $7.48^{+0.82}_{-0.86} \times 10^{-5}$ \\
\hline
\multirow{2}{*}{$\mu (n_B=0.05)$}  & $5.6^{+2.3}_{-1.8} \times 10^{-4}$ &  $4.6^{+2.2}_{-2.4} \times 10^{-4}$ \\
                                    & $5.78^{+0.69}_{-0.75} \times 10^{-4}$ & $5.49^{+0.66}_{-0.72} \times 10^{-4}$ \\
\hline
\multirow{2}{*}{$f=\omega_0/(2\pi)$} & $38.6^{+4.7}_{-4.0}$ &  $33.7^{+2.7}_{-3.5}$\\
                                    & $42.9^{+1.5}_{-1.8}$ & $35.09^{+0.92}_{-0.96}$ \\
\hline
\end{tabular}
\tablefoot{The shear modulus (in MeV~fm$^{-3}$) is given at three different chosen densities in the crust (indicated in parenthesis, in fm$^{-3}$; top three rows), while the frequency (in Hz) of the $_2t_0$ mode is given for a $M=1.4 M_{\odot}$ NS (last row). Computations either neglect (labelled `w/o finite size') or include (labelled `with finite size') the finite-size correction in Eq.~\eqref{eq:shear-mod-corr}, and assume entrainment according to \citet{Almirante2025}, that is, $\rho_{B,{\rm d}}=\rho_B - \rho_{B,{\rm g}}$ in Eq.~\eqref{eq:shear-speed}. For each quantity, the first (second) row corresponds to the calculation starting from a uniform (nuclear-physics-informed) prior (points i and ii, respectively, in Sect.~\ref{sec:bayes}).
}
\label{tab:median}
\end{table}

\end{appendix}

\end{document}